\documentclass[aps,prc,reprint,superscriptaddress,showpacs,amsmath,amssymb,longbibliography,lengthcheck]{revtex4-1}
\usepackage{graphicx}
\usepackage{txfonts}

\begin{document}

\title{Search for three alpha states around an $^{16}$O core in $^{28}$Si}

\author{T.~Ichikawa}
\affiliation{Yukawa Institute for Theoretical Physics, Kyoto University,
Kyoto 606-8502, Japan}

\author{N.~Itagaki}
\affiliation{Yukawa Institute for Theoretical Physics, Kyoto University,
Kyoto 606-8502, Japan}

\author{Y.~Kanada-En'yo}
\affiliation{Department of Physics, Kyoto University, Kyoto 606-8502, Japan}

\author{Tz.~Kokalova}
\affiliation{
School of Physics and Astronomy, University
of Birmingham, Edgbaston, B15 2TT, Birmingham, UK}

\author{W.~von~Oertzen}
\affiliation{
Helmholtz-Zentrum Berlin, Hahn-Meitner-Platz 1, 14109 Berlin, Germany}

\date{\today}

\begin{abstract}
 We investigate the existence of weakly coupled gas-like states
 comprised of three $\alpha$ particles around an $^{16}$O core in
 $^{28}$Si.  We calculate the excited states in $^{28}$Si using the
 multi-configuration mixing method based on the $^{16}$O + 3$\alpha$
 cluster model. We also include the $^{16}$O + $^{12}$C and $^{24}$Mg +
 $\alpha$ basis wave functions prepared by the generator coordinate
 method. To identify the gas-like states, we calculate the isoscalar
 monopole transition strengths and the overlap of the obtained states
 with the geometrical cluster wave function and the
 Tohsaki-Horiuchi-Schuck-R\"{o}pke (THSR) wave function.  The results
 show that the obtained fourth and twelfth states significantly overlap
 with the THSR wave function. These two states clearly coexist with the
 $^{16}$O + $^{12}$C cluster states, emerging at similar energies. The
 calculated isoscalar monopole strengths between those two states are
 significantly large, indicating that the states are members of the
 excitation mode. Furthermore, the calculated root-mean-squared (RMS)
 radii for these states also suggest that a layer of gas-like three
 $\alpha$ particles could exist around the surface of the $^{16}$O core,
 which can be described as a ``two-dimensional gas'' in the intermediate
 state before the Hoyle-like three $\alpha$ states emerge.
\end{abstract}

\maketitle

The investigations of excited states in light-mass nuclei provide a good
opportunity to study the rich variety of nuclear structures of quantum
many-body systems. One of these states is the well known Hoyle state in
$^{12}$C$^*$ \cite{Ho54}. The Hoyle state is the second 0$^+$ state in
$^{12}$C$^*$, with an excitation energy of 7.65 MeV, just above the 3
$\alpha$ decay threshold energy. Many theoretical attempts have been
performed to reproduce its energy and geometrical properties. It was
found, that it is difficult to reproduce those values by calculations
based only on the shell model. On the other hand, the microscopic three
$\alpha$ cluster models successfully describe the properties of the
Hoyle state (such as the observed $\alpha$-decay width) and indicate
that the three $\alpha$ state develops well in the $0^+_2$ state
\cite{Uegaki,Kamimura}.

Recent calculations, using the $\alpha$ cluster model suggest that in
the Hoyle state, the three ``gas-like'' $\alpha$ particles are weakly
coupled with each other at near the three $\alpha$-decay threshold
energy \cite{Tohsaki,Schuck}.  Based on this picture, the
Tohsaki-Horiuchi-Schuck-R\"{o}pke (THSR) wave function was proposed.
This wave function describes the $\alpha$ particles as independent, with
their center-of-mass motion, in the same $0S$ state of the harmonic
oscillator, having a large oscillator parameter.  Using this wave
function, Tohsaki {\it et al.}  proposed the extremely large RMS nuclear
radius for the $0^+_2$ state in $^{12}$C \cite{Tohsaki}.

Those studies have triggered interest in whether such gas-like states
exist in heavier-mass nuclei.  Funaki {\it et al.} searched for gas-like
four $\alpha$ states in $^{16}$O$^*$ using the microscopic $\alpha$
cluster model coupled with the THSR wave function
\cite{Funaki,Funaki03}.  In this connection, the existence of weakly
coupled gas-like $\alpha$ cluster states around a {\it core} has been
recently suggested \cite{Kokalova-1,Kokalova-2,vOe,vOe1,Ogloblin}. To
study this, the Monte-Carlo technique for the THSR wave function was
proposed \cite{VT}.  The possibility of a gas-like three $\alpha$
cluster state around $^{40}$Ca in $^{52}$Fe has been discussed using
this technique \cite{40Ca}.  To identify such gas-like states
experimentally, the measurement of an isoscalar monopole transition
strength has been proposed \cite{Yamada,Kawabata}.  The enhancement of
the transition strength would correspond to the development of such
cluster structure.  In this respect, gas-like two $\alpha$ states around
an $^{16}$O core in $^{24}$Mg have been investigated both theoretically
\cite{Yamada,Ichikawa,YIO,Peru} and experimentally
\cite{Kawabata,Wakasa,Sasamoto}.
  
The aim of the present paper is to search for the weakly coupled
gas-like three $\alpha$ cluster states around an $^{16}$O core in
$^{28}$Si.  In analogy to the three $\alpha$-particle Hoyle state in
$^{12}$C, the existence of states, similar to the Hoyle state with three
$\alpha$-particles around an $^{16}$O core can be expected.  However, in
$^{28}$Si, many strongly coupled cluster states would emerge at and
below/above the ($^{16}$O + 3$\alpha$) decay threshold energy
\cite{Hori10,enyo10,ichi10}.  Three $\alpha$ particles around an
$^{16}$O core would form the strongly coupled $^{12}$C state, to gain
the inter cluster potential energy.  An important feature is that these
weakly and strongly coupled systems would coexist and that their
emergencies compete with each other. To discuss the existence of the
weakly coupled gas-like three $\alpha$ state around an $^{16}$O core, it
is important to completely understand the occurrence of all strongly
coupled cluster states within the orthogonality condition among the
states.

To investigate the existence of the gas-like three $\alpha$ state around
the $^{16}$O core, we calculate the excited states of $^{28}$Si using
the $^{16}$O + 3$\alpha$ cluster model.  We superpose many randomly
generated Slater determinants, which correspond to the
multi-configuration mixing (MCM) method. To describe the molecular
states, we also include the basis wave functions with $^{16}$O +
$^{12}$C and $^{24}$Mg + $\alpha$ configurations prepared by the
generator coordinate method (GCM).  In the MCM calculations, it becomes
often difficult to identify the structure of the obtained results as the
number of the basis wave functions increases, because many states couple
with each other. Thus, we must develop the analyzing method for the
results from the full large scale calculations.  To identify the
relevant states, we calculate the overlap of the obtained results with
the THSR and the geometrical cluster wave functions. We also calculate
the RMS radius and the isoscalar monopole transition strength.  We show
below how the gas-like three $\alpha$ states emerge and how they connect
with the ground state.

\begin{figure}
\includegraphics[width=\linewidth]{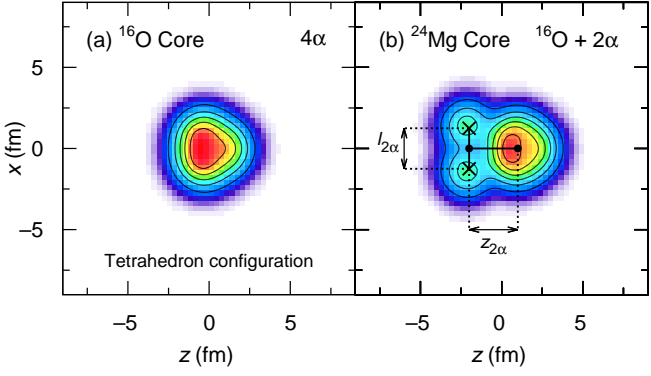} \caption{(Color online) Density
 plots of the basis wave functions for the $^{16}$O and $^{24}$Mg cores. The
 contours correspond to multiple steps of 0.2 fm$^{-2}$. The color is
 normalized by the largest density in each plot. }
\label{fig1}
\end{figure}

In the present study, we use Brink's $\alpha$-cluster
model~\cite{brink}.  In this approach the wave function for $^{28}$Si is
described by the Slater determinant of the seven $\alpha$ clusters. The
wave function of the $\alpha$ cluster, $\phi_\alpha$, is described by
the direct product of four wave functions for each nucleon (proton and
neutron with spin down and up). The spacial part of the wave function
for the $\alpha$ cluster is given by $\phi_\alpha({\vec
R})\propto\prod_{i=1}^4\exp[-\nu(\vec r_i-\vec R)^2]$, where $\vec R$
denotes the center position vector of the Gaussian function, $\vec r_i$
denotes the spatial coordinate of each nucleon, and $\nu$ denotes the
size parameter. We take $\nu=1/2b^2$ with $b=1.46$ fm in the
calculations. To describe the $\alpha$ cluster, we take the same $\vec
R$ for all four nucleons.

To calculate the excited states, we superimpose many basis wave
functions described by the Brink $\alpha$-cluster model for
$^{28}$Si. In this study, we prepare the basis wave functions using
$^{16}$O + 3$\alpha$ cluster configurations to investigate the three
$\alpha$ states around the $^{16}$O core. To take into account the
molecular states, we also prepare the basis wave functions using
$^{16}$O + $^{12}$C and $^{24}$Mg + $\alpha$ configurations based on the
GCM.

We first prepare the cluster wave functions using the $^{16}$O,
$^{12}$C, and $^{24}$Mg configurations. The cluster wave functions of
$^{16}$O, $^{12}$C, and $^{24}$Mg with the center-of-mass position $\vec
R$ are denoted by $\Phi_{^{16}{\rm O}}(\vec R)$, $\Phi_{^{12}{\rm
C}}(\vec R)$, and $\Phi_{^{24}\rm Mg}(\vec R)$, respectively. For
$\Phi_{^{16}{\rm O}}(\vec R)$, we take a tetrahedron configuration with
four $\alpha$ clusters. The additional three $\alpha$'s form a triangle,
it is perpendicular to the $z$-axis and one side of the triangle is
parallel to the $x$-$z$ plane. The further residual $\alpha$ particle is
placed on the positive $z$ direction. Figure 1(a) shows the total
density of $\Phi_{^{16}{\rm O}}$ used in this study. The
$\alpha$-$\alpha$ distance between each center position of the Gaussian
functions of the four $\alpha$'s in the $^{16}$O core is quite small,
0.5 fm, in order to simulate the doubly closed configuration of the
shell model.

For $\Phi_{^{12}{\rm C}}(\vec R)$, we take the equilateral triangle
configuration of the three $\alpha$ clusters with a length of $a$ fm.
The triangle is placed on the $x$-$z$ plane and a side (two of the
$\alpha$ clusters) is perpendicular to the $z$ axis in the
center-of-mass flame of $^{12}$C. The residual
$\alpha$ cluster is placed on the negative $z$ axis. We also take three
different configurations by rotating the $\alpha$-cluster triangle on
the $x$ axis with angles of 0$^\circ$, 45$^{\circ}$, and
90$^{\circ}$. Those configurations are denoted by
$^{12}$C$_{\parallel}$, $^{12}$C$_{\angle}$ and $^{12}$C$_{\perp}$.

The wave function of $\Phi_{^{24}\rm Mg}$($\vec R$) is an isosceles
configuration with $\Phi_{^{16}\rm O}$($\vec R_0$) and two $\alpha$
clusters.  Figure 1(b) shows the density plot of $\Phi_{^{24}\rm Mg}$
used in this study. The configuration of $\Phi_{^{16}\rm O}$($\vec R_0$)
is the same as mentioned above. The isosceles triangle is placed on the
$x$-$z$ plane and the base of the two alpha clusters is perpendicular to
the $z$ axis.  The crosses denote the positions of the two $\alpha$
clusters. The distance between the two $\alpha$ clusters is denoted by
$l_{2\alpha}$. The distance between $\Phi_{^{16}\rm O}$($\vec R_0$) and
the $z$ position of the base is denoted by $z_{2\alpha}$.  We determine
$l_{2\alpha}$ and $z_{2\alpha}$ by minimizing the expectation value of
the total energy for $\Phi_{^{24}{\rm Mg}}$.  We here take
$l_{2\alpha}=2.5$ fm and $z_{2\alpha}=3.0$ fm.

We next prepare the basis wave functions using $^{16}$O + 3$\alpha$,
$^{16}$O + $^{12}$C, and $^{24}$Mg + $\alpha$ cluster configurations.
The basis wave function using the $^{16}$O + 3$\alpha$ cluster
configuration, $\Psi^{(i)}_{^{16}{\rm O} + 3\alpha}$, is given by
$ \Psi^{(i)}_{^{16}{\rm O} + 3\alpha}= [ {\cal
  A}\Phi_{^{16}{\rm O}}(\vec R_0)\phi_\alpha(\vec R_1)\phi_{\alpha}(\vec
  R_2)\phi_{\alpha}(\vec R_3) ]_i$, 
where ${\cal A}$ denotes the anti-symmetrization operator.  We first
take $\vec R_0=0$ and the positions of the three $\alpha$ clusters $\vec
R_1$, $\vec R_2$, and $\vec R_3$ are randomly generated under the
condition of the distributed probability $w(\vec R_i)$.  In this study,
we use $w(\vec R_i)\propto\exp[-{\vec R^2_i}/\sigma^2]$, where $\sigma$
is the size parameter.  Subsequently, the center-of-mass correction for
$ \Psi^{(i)}_{^{16}{\rm O} + 3\alpha}$ was performed.

The basis wave functions using the $^{16}$O + $^{12}$C configuration,
$\Psi^{(i)}_{^{16}{\rm O}+^{12}{\rm C}}$, is given by
$\Psi^{(i)}_{^{16}{\rm O} + ^{12}{\rm C}}= [{\cal A}\Phi_{^{16}{\rm
O}}(\vec R_0)\Phi_{^{12}{\rm C}}(\vec R)]_i$.  The $^{12}$C cluster is
shifted to the positive $z$ direction. Here, the distance between the
center-of-mass positions of the $^{16}$O and $^{12}$C clusters is
denoted by $d_{^{16}{\rm O\mathchar`-}^{12}{\rm C}}$.  The basis wave functions
using $^{24}$Mg + $\alpha$ configurations, $\Psi^{(i)}_{^{24}{\rm Mg} +
\alpha}$, is given by $\Psi^{(i)}_{^{24}{\rm Mg} + \alpha}= [{\cal
A}\Phi_{^{24}{\rm Mg}}(\vec R_0)\phi_\alpha(\vec R)]_i$.  The wave
function of $\phi_\alpha(\vec R)$ is placed on the $x$, $y$, or $z$
axes. Those configurations are denoted by $^{24}$Mg + $\alpha_x$,
$^{24}$Mg + $\alpha_y$, and $^{24}$Mg + $\alpha_z$. The distance between
the center-of-mass positions of the $^{24}$Mg and $\alpha$ clusters is
denoted by $d_{^{24}{\rm Mg\mathchar`-}\alpha}$.

In the calculation, we generate 1000 wave functions for $\Psi_{^{16}{\rm
O} + 3\alpha}^{(i)}$.  For $\Psi_{^{16}{\rm O}+^{12}{\rm C}}^{(i)}$, we
take four and eight values for $a$ and $d_{^{16}{\rm O\mathchar`-}^{12}{\rm C}}$
in each angle of $^{12}$C$_{\parallel}$, $^{12}$C$_{\angle}$ and
$^{12}$C$_{\perp}$, respectively. Those correspond to $a=(1.0, 2.0, 3.0,
4.0)$ fm and $d_{^{16}{\rm O\mathchar`-}^{12}{\rm C}}=(1.0, 2.0, \dots, 8.0)$ fm.
For $\Psi_{^{24}{\rm Mg} + \alpha}^{(i)}$, we take eight values for
$d_{^{24}{\rm Mg\mathchar`-}\alpha}$ corresponding to $d_{^{24}{\rm
Mg\mathchar`-}\alpha}=(1.0, 2.0, \dots, 8.0)$ fm in each direction of $\alpha_x$,
$\alpha_y$, and $\alpha_z$.

Thus, the total number of the basis wave functions is 1120.  The total
wave function $\Psi_{\rm Total}$ is described with the linear
combination of the basis wave functions given by $\Psi_{\rm
Total}=\sum_i c_i P^+ P^0_{00}\Psi^{(i)}$, where $\Psi^{(i)}$ denotes
each basis wave function.  The coefficient $c_i$ is determined by
diagonalizing the total Hamiltonian.  The symbols $P^+$ and $P^0_{00}$
denote the projection operator for the parity and angular-momentum
($J=M=K=0$) to the 0$^+$ state, respectively.  We here only calculate
the 0$^{+}$ states to focus on the gas-like three $\alpha$ state. We
perform the center-of-mass correction and the parity and angular
momentum projections to the $0^{+}$ state for each basis wave function
numerically. For the calculations of the angular-momentum projection,
$24\times32\times24$ grids points are taken with respect to the
$\alpha$, $\beta$, and $\gamma$ directions of the Euler angle.  For the
total Hamiltonian, we use the Volkov No.2 effective $N$-$N$
interaction~\cite{Vol}.  In the calculation, the Majorana exchange
parameter, $M$, is chosen to be 0.645 so as to reproduce the binding
energy of $^{28}$Si.  In the calculation, the cluster decay threshold
energies for $^{16}$O + 3$\alpha$ and $^{16}$O + $^{12}$C are $-189.29$
MeV and $-186.48$, respectively.

\begin{figure*}
\includegraphics[width=16.8cm]{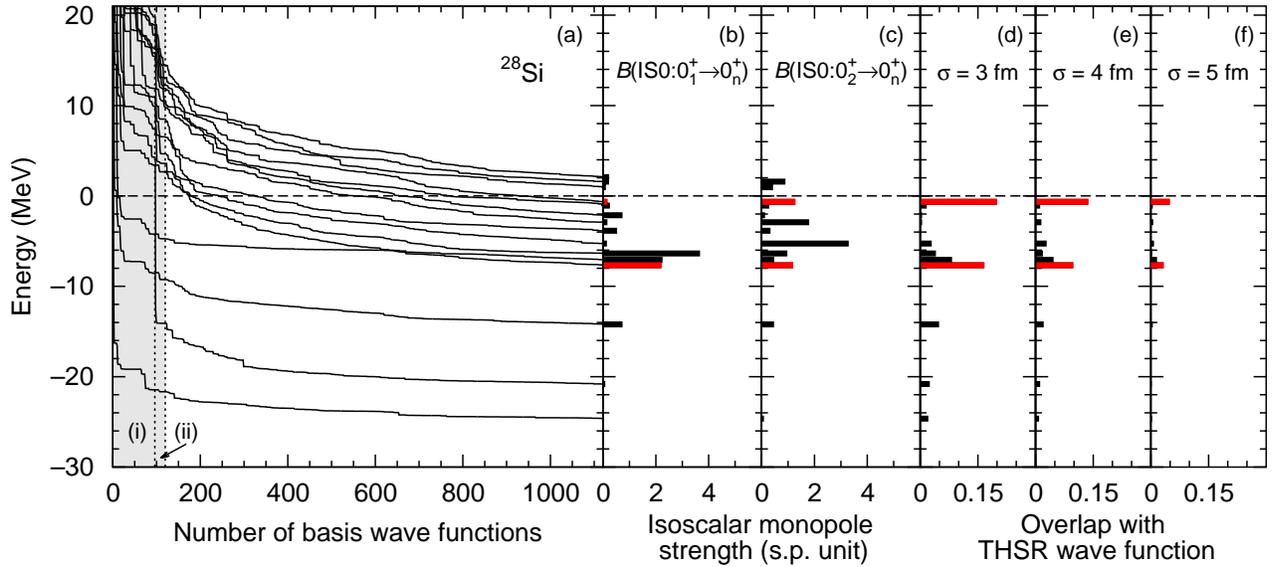} \caption{(Color online) (a)
Convergence behavior of the calculated energies versus the number of
basis wave functions measured from the $^{16}$O + 3$\alpha$ decay
threshold energy. In the gray area denoted by (i) and (ii), we use the
basis wave functions with $^{16}$O + $^{12}$C and $^{24}$Mg + $\alpha$
configurations, respectively. (b) and (c) Isoscalar monopole-transition
strength from the first and second states in the unit of the
single-particle (S.P.) excitation from the 1$s$ to the 2$s$ states,
respectively. (d), (e), and (f) Overlap between each state and the THSR
wave function with $\sigma=$ 3, 4, and 5 fm, respectively.}
\label{fig2}
\end{figure*}


To identify the gas-like $\alpha$ states, we calculate the overlap
between the obtained states and the THSR wave function.  The THSR wave
function with the gas-like three $\alpha$ particles around an $^{16}$O
core, $\Psi^{(\sigma)}_{\rm THSR}$, is given by
\begin{eqnarray}
  \Psi^{(\sigma)}_{\rm THSR} = && \int d \vec{R_1} d\vec{R_2}d\vec{R_3}
   \cdot \exp[ -(\vec R_1^2 + \vec R_2^2 + \vec R_3^2)/\sigma^2 ] \nonumber \\
&&\cdot [{\cal A}  \phi_{^{16}{\rm O}}(\vec R_0)\phi_\alpha(\vec R_1)\phi_\alpha(\vec R_2)\phi_\alpha(\vec R_3) ] \nonumber \\
= && {\cal A}\phi_{^{16}{\rm O}}(\vec R_0) \prod_{i=1}^3 \int d \vec R_i \ \phi_\alpha(\vec R_i) \exp[-\vec R_i^2 / \sigma^2],
\end{eqnarray}

where $\sigma$ is the size parameter of the gas-like $\alpha$ state. We
can perform the Monte-Carlo integration for Eq.~1, when the exponential
term $\exp[-\vec R_i^2 / \sigma^2]$ in Eq.~1 is regarded as the weight
factor of the randomly generated integral points. We also perform the
parity and angular-momentum projections to the $0^+$ state at each
generated integral point.  Then, the Monte-Carlo THSR wave function
$\Psi^{(\sigma)}_{\rm THSR}$ is given by $\Psi^{(\sigma)}_{\rm
THSR}=\sum_iP^+ P^0_{00}\Psi^{(i)}_{^{16}{\rm O}+3\alpha}$. The integral
points for $\vec R_1$, $\vec R_2$, and $\vec R_3$ are randomly generated
under the condition of the weight factor $w({\vec R_i})$ given by
$w({\vec R_i})\propto \exp[-\vec R_i^2 / \sigma^2]$.  In this study, we
calculate the THSR wave function with $\sigma=$ 3, 4 and 5 fm.  For the
Monte-Carlo calculations, we use 900 randomly generated basis wave
functions for each $\sigma$. The expectation values of the total energy
for those with $\sigma=$ 3, 4 and 5 converge well at energies of
$-191.39$, $-188.48$ and $-182.33$ MeV, respectively.

To investigate the properties of the states obtained in this procedure,
we also calculate the isoscalar monopole-transition strength and the RMS
radius.  Here, the unit of the transition strength, $B_0({\rm IS0})$, is
taken as the value obtained from the 1$s$ to the 2$s$ states, described
by the single-particle wave function of the three-dimensional harmonic
oscillator. The value of $B_0({\rm IS0})$ is then given by $B_0({\rm
IS0})=\sqrt{5}b^2$ fm$^2$ \cite{Yamada}. In the present study, we take
$B_0({\rm IS0})=4.77$ fm$^{2}$ with $b=1.46$.


Figure \ref{fig2} (a) shows the convergence behavior of the states
measured from the $^{16}$O + 3$\alpha$ decay threshold energy versus the
number of the basis wave functions. We plot here only the results below
the 15th state. In Fig.~2(a), the gray area (i) denotes the results
calculated with the subspace for the $^{16}$O + $^{12}$C$_{\perp}$,
$^{12}$C$_{\parallel}$, and $^{12}$C$_{\angle}$ configurations.  The
gray area (ii) also denotes those for the $^{24}$Mg + $\alpha_x$,
$\alpha_y$, and $\alpha_z$ configurations in addition to the $^{16}$O +
$^{12}$C components. The lowest energy in the calculated states with the
total wave functions is $-24.64$ MeV, which is in good agreement with
the experimental value of $-24.03$ MeV.

\begin{figure}
\includegraphics[width=\linewidth]{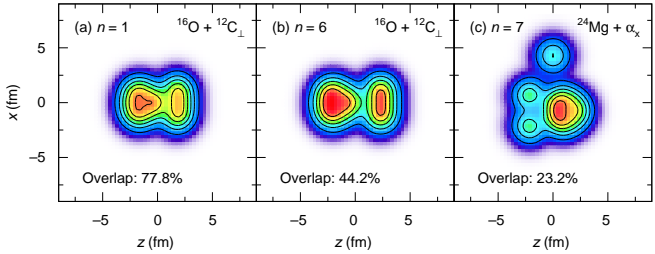}
\caption{(Color online) Density plots of the basis wave function
 overlapping maximally with the calculated states. The contours 
 correspond to multiples of 0.2 fm$^{-2}$. The color is normalized by
 the largest density in each plot.}
\label{fig3}
\end{figure}

We first investigate the structure of the ground (first) state.  This
state corresponds to the prolate shape with a significant $^{16}$O +
$^{12}$C$_\perp$ cluster component, which originates from the lowest
state of the results of the gray area (i) in Fig.~2(a). The RMS radius
obtained for this state is 2.80 fm. To clarify the cluster structure,
the overlap of the obtained states with the wave functions using
$^{16}$O + $^{12}$C and $^{24}$Mg + $\alpha$ configuration is
calculated.  Figure~\ref{fig3} shows the density plots of the wave
functions maximally overlapping with the calculated states. The lowest
state overlaps maximally with the $^{16}$O + $^{12}$C$_\perp$
configuration of 78.5\% at $a=2.0$~fm and $d_{^{16}{\rm O\mathchar`-}^{12}{\rm
C}}=3.0$~fm (see Fig.~3(a)).

The second state corresponds to the one-body oblate shape. The obtained
RMS radius is 2.77 fm, which is somewhat smaller than that of the first
state. The second state overlaps maximally with the $^{16}$O +
$^{12}$C$_\parallel$ configuration of 17.3\% at $a=3.0$ fm and
$d_{^{16}{\rm O\mathchar`-}^{12}{\rm C}}=1.0$ fm, but many other components also
have similar overlaps at small $d_{^{16}{\rm O\mathchar`-}^{12}{\rm C}}$. Thus,
this state is the compact one-body system rather than a specific cluster
state. In Brink's $\alpha$-cluster model without the spin-orbit force,
the shape configurations for the first and second $0^+$ states are
prolate and oblate (pentagon), respectively. This shows that the results
of our calculations are consistent with previous studies, although the
experimental data suggest that the ground and the third $0^+$ states
correspond to the oblate and the prolate shapes, respectively
\cite{Bauhoff}.

We next investigate the structure of the excited states. Figure 2(b) and
(c) show the calculated isoscalar monopole-transition strength from the
first and second states in the unit of $B_0({\rm IS0})$, respectively.
Figure 2(d), (e), and (f) show the calculated overlaps between the
obtained states and the THSR wave function with $\sigma=3$, 4, and 5 fm,
respectively. Below, we only discuss the characteristic results which
can be identified by the analysis.

In Figure \ref{fig2} (d), (e), and (f), we can see that the fourth and
twelfth states significantly overlap with the THSR wave function. These
states are candidates for the gas-like three $\alpha$ state discussed in
this study. The overlap between the fourth state and the THSR wave
functions with $\sigma=$3, 4, and 5 fm are 16.7\%, 9.8\% and 3.3\%,
respectively. Those overlaps for the twelfth state are 20.0\%, 13.7\%
and 4.9\%.  We also orthogonalize the THSR wave functions with $\sigma=$
3, 4, and 5 fm and calculate the overlaps, but the results are very
similar to the largest value with the single THSR wave function.  The
RMS radii obtained for the fourth and the twelfth states are 2.92 and
3.02 fm, respectively. These values are somewhat spatially extended
rather than that of the first and second states.

In addition, we obtain remarkably strong transition strength from the
fourth to twelfth states by $B({\rm IS0})/B_0({\rm IS0})=3.20$,
indicating that those states strongly correlate with each other due to
the cluster excitation.  We thus consider that the fourth and twelfth
states are a member of the new excitation mode.  We also obtain a
relatively large transition strength from the first to the fourth states
by $B({\rm IS0})/B_0({\rm IS0})=2.21$ (see Fig.~2 (b)), suggesting that
this new excitation mode is built on the first state with the prolate
shape.

The sixth state has the strongest transition strength from the ground
state by $B({\rm IS0})$/$B_0({\rm IS0})=3.67$. This state overlaps
maximally with the $^{16}$O + $^{12}$C$_\perp$ configuration of 42.2\%
at $a=2.0$~fm and $d_{^{16}{\rm O\mathchar`-}^{12}{\rm C}}=4.0$ fm (see
Fig.~3(b)). The obtained RMS radius is 2.95 fm.  This corresponds to the
$^{16}$O + $^{12}$C molecular state where the relative motion between
two clusters is excited from the ground (first) to the higher-nodal
states.  The seventh state also has strong transition strength from the
second state by $B({\rm IS0})$/$B_0({\rm IS0})=3.30$. This state is the
cluster excitation of the $^{24}$Mg + $\alpha_x$ configuration. The
maximum overlap between the seventh state and the $^{24}$Mg + $\alpha_x$
configuration is 23.2\% at $d_{^{24}{\rm Mg\mathchar`-}\alpha}=5.0$ fm
(see Fig.~3(c)). The obtained RMS radius is 2.93~fm.

An important observation is that the fourth and the twelfth states do
not largely overlap with any geometrical configurations of the $\alpha$
clusters. Despite this, the RMS radii for these states are considerably
extended to similar extent as other cluster states. The fourth and
twelfth excited states significantly overlap with the THSR wave function
not only with $\sigma=3$ fm but also $\sigma=4$ fm. This suggests the
existence of gas-like three $\alpha$ particles spreading to the outside
of the $^{16}$O core.  To see the extent of the spreading for the
gas-like three $\alpha$ particles, we estimate the RMS radius of the
three $\alpha$ particles. Here, the effect of the $^{16}$O core is
subtracted by neglecting the recoil correction.  The RMS radius for the
three $\alpha$ particles, $r_{\rm RMS}^{(3\alpha)}$, is then given by
$r_{\rm RMS}^{(3\alpha)}=\sqrt{(28<r^2(^{28}{\rm Si})> -
16<r^2(^{16}{\rm O})>)/12}$, where $<r^2(^{28}{\rm Si})>$ and
$<r^2(^{16}{\rm O})>$ denote the expectation value of the squared radius
for the $^{28}$Si and the $^{16}$O core, respectively. We calculate
$<r^2(^{28}{\rm Si})>$ without the center-of-mass correction. We also
obtain $\sqrt{<r^2(^{16}{\rm O})>}=2.2$ fm with the tetrahedron
configuration of four $\alpha$'s.  The calculated RMS radii of the three
$\alpha$ particles for the first, fourth, and twelfth states are 3.45,
3.67, and 3.86 fm, respectively.  Those are indeed expanded with
increasing excitation energies. However, the spatial distributions for
those three $\alpha$ particles are not so widely spread rather than that
of the Hoyle state in $^{12}$C because of the existence of the
attractive $^{16}$O core.

Therefore, we consider that the fourth and twelfth states are members of
a new type of excitation mode built on the prolate state, which has not
yet been well established in experiments. In the first excited state of
this mode (fourth state), the weakly coupled gas-like three $\alpha$
particles, without the angular correlations, emerge around the surface
of the $^{16}$O core. That can be described as a layer of dilute density
formed around a core, which would be regarded as a ``two-dimensional
(2D) gas''.  The emergence of this weakly coupled 2D gas-like state
significantly competes with that of the $^{16}$O + $^{12}$C molecular
state (sixth state), which have strongly coupled three $\alpha$'s of
$^{12}$C, as 
their energies are almost comparable to each other.  This indicates,
that the energy necessary for the excitation from the strongly coupled
$^{12}$C to the weakly coupled three $\alpha$'s around the $^{16}$O core
is comparable to the relative motion between $^{16}$O and $^{12}$C.
Despite this competition, the coexistence of the 2D gas-like and the
strongly coupled states is highly possible, as shown in this study.

In the second excited state of this new mode (twelfth state), the
gas-like layer is well developed and more spatially expanded, because 
the isoscalar monopole strength only from the first excited (fourth)
state is extremely large.  In more highly excited states, the Hoyle-like
weakly coupled 3$\alpha$ state, that is a ``three-dimensional (3D)
gas-like'' state around the $^{16}$O core may emerge. Then, the second
excited 2D gas-like state would be the intermediate state before the 3D
gas-like state emerges and would be directly connected to the 3D
gas-like state by the large isoscalar monopole transition
strength. However, it is difficult to identify such 3D gas-like states
in the present calculations, because many continuum states are coupled
with other states in the high excitation energies.  The development of
analyses, such as the complex scaling and the pseudo-potential method,
is necessary when searching for such states.


In summary, we have investigated the existence of weakly coupled
gas-like three $\alpha$ states around an $^{16}$O core. To study this,
we have calculated the excited states of $^{28}$Si using the
multi-configuration mixing method with the basis wave functions randomly
generated by the $^{16}$O + 3$\alpha$ cluster model. We have also
included the $^{16}$O + $^{12}$C and $^{24}$Mg + $\alpha$ basis wave
functions, prepared by the GCM method to describe well the molecular
states. To identify the states, which we have obtained, we have
calculated their overlap with the geometrical cluster wave functions and
the THSR wave function. Furthermore we have also calculated the RMS
radius and the isoscalar monopole transition strength for the obtained
states.

We have found that the fourth and twelfth excited states largely overlap
with the THSR wave function with $\sigma=$3 and 4 fm. The calculated
isoscalar monopole strengths are also significantly large from the first
to the fourth and from the fourth to the twelfth states, indicating
those may be members of a new excitation mode. At around the energy of
the fourth excited state, the $^{16}$O + $^{12}$C cluster state, which
has three strongly coupled $\alpha$'s of $^{12}$C, also emerges, but the
fourth state coexists with this state. The calculated RMS radii for the
fourth and twelfth states suggests that a layer of dilute three $\alpha$
particles may exist around the $^{16}$O core.  This gas-like structure
in the layer would be the intermediate state, before the
complete 3D Hoyle-like gas state emerges.

\begin{acknowledgments}
A part of this research has been funded by MEXT HPCI STRATEGIC PROGRAM.
TzK is grateful to the Daphne Jackson trust and STFC for their support.
Numerical computation in this work has been carried out at the Yukawa
Institute Computer Facility using the SR16000 system. This work was
performed as part of the Yukawa International Project for Quark-Hadron
Sciences (YIPQS). It was also supported by a Grant-in-Aid for the Global
COE Program ``The Next Generation of Physics, Spun from Universality and
Emergence'' from the Ministry of Education, Culture, Sports, Science and
Technology (MEXT) of Japan and a Grant-in-Aid for Scientific Research
from the Japan Society for the Promotion of Science.
\end{acknowledgments}

\end{document}